\documentclass[useAMS,usenatbib]{mn2e}
\pdfoutput=1
\usepackage{graphicx}

\newcommand{\ml}{M/L}
\newcommand{\mlv}{M/L_\mathrm{V}}
\newcommand{\mlk}{M/L_\mathrm{K}}

\title[The local stellar luminosity function and mass-to-light ratio in the NIR]
{The local stellar luminosity function and mass-to-light ratio in the NIR}

\author[A. Just, B. Fuchs, H. Jahrei{\ss}, C. Flynn, C. Dettbarn, J. Rybizki]
{
A. Just$^{1}$\thanks{E-mail:just@ari.uni-heidelberg.de},
B. Fuchs$^{1}$, 
H. Jahrei{\ss}$^{1}$, 
C. Flynn$^{2}$,
C. Dettbarn$^{1}$,
J. Rybizki$^{1}$\\
$^{1}$Astronomisches Rechen-Institut, Zentrum f\"ur Astronomie der
Universit\"at Heidelberg (ZAH),\\
M\"onchhofstra{\ss}e 12-14, 69120, Heidelberg, Germany\\
$^{2}$Centre for Astrophysics and Supercomputing, Swinburne University, \\
Hawthorn 3122, Melbourne, Australia
\\
}

\begin{document}

\maketitle

\label{firstpage}

\begin{abstract}
A new sample of stars, representative of the solar neighbourhood luminosity
function, is constructed from the {\it Hipparcos} catalogue and the Fifth Catalogue
of Nearby Stars. We have cross-matched to sources in the 2MASS catalogue so
that for all stars individually determined Near Infrared photometry (NIR) is
available on a homogeneous system (typically $K_{\rm s}$). The spatial
completeness of the sample has been carefully determined by statistical
methods, and the NIR luminosity function of the stars has been derived by
direct star counts.
We find a local volume luminosity of $0.121\pm 0.004\, L_{K\sun}\mathrm{pc}^{-3}$, 
corresponding to a volumetric mass-to-light ratio of $M/L_K =
0.31 \pm 0.02\,{\mathcal{M}}_{\sun}/L_{K\sun}$, where
giants contribute 80 per cent to the light but less than 2 per cent to the stellar 
mass. We derive the surface brightness of the solar cylinder with the help of a 
vertical disc model. We find a surface brightness 
of $99\, L_{K\sun}\mathrm{pc}^{-2}$ with an uncertainty of approximately 
10 per cent. This corresponds to a mass-to-light ratio for 
the solar cylinder of $M/L_K = 0.34\,{\mathcal{M}}_{\sun}/L_{K\sun}$. 
The mass-to-light ratio for the solar cylinder is only 10 per cent larger than the 
local value despite the fact that the local population has a much larger contribution 
of young stars. It turns out that the effective scale heights of the lower main 
sequence carrying most of the mass is similar to that of the giants, which are 
dominating the NIR light. The corresponding colour for 
the solar cylinder is $V-K=2.89$\,mag compared to the local value of $V-K=2.46$\,mag.
An extrapolation of the local surface brightness to the whole Milky Way yields a 
total luminosity of $M_K=-24.2$\,mag. The Milky Way falls in the range of K band 
Tully-Fisher (TF) relations from the literature.

\end{abstract}


% --------------------------------------------------- Intro ----- Section 1
\section{Introduction}

The stellar luminosity function (hereafter LF), i.e.\ the inventory of stars as
a function of their absolute magnitudes, is a fundamental property of a stellar
population, and has wide implications for understanding star formation, and the
formation and evolution of galaxies. At present, we can determine a complete LF
on a star-by-star basis, down to the hydrogen burning limit, for stars in the
Milky Way only. Such LFs have been determined for stars in the solar
neighbourhood, and also for open clusters, globular clusters and in the Galactic
bulge. Conventionally, the LF refers to luminosities of the stars in the
optical bands, such as the $V$\,band. One of the most influential determinations
of the nearby LF is that of \citet{Wie74}, who based the study on the
'Catalogue of Nearby Stars' \citep[]{Gli}, an extensive compilation of data
on all stars within 22\,pc of the Sun. More recently, \citet{Fly06} 
have re-determined the LF using modern data, 
including in particular the {\it Hipparcos}
data \citep[]{hipcat}, but finding only small corrections to the classic work. Those
authors determined not only local star number densities \citep[as in][]{Wie74}, but
also derived surface densities of stars (i.e.\ in a column integrated through
the Galactic disc) as a function of absolute magnitude. In surface density
terms, light from the Milky Way disc was found to be dominated by emission from
main sequence stars with $M_V \approx 1$\,mag (spectral type A) and by old K-
and M-giants with $M_V \ga 1$\,mag \citep{Hou78}.

For the analysis of the dynamics and evolution of galaxies, it is necessary to
determine the masses of the Galactic components. From
population synthesis it is obvious that the mass-to-light ratio (hereafter $\ml$, 
in solar units) in the optical is very
sensitive to the contribution of very young stellar populations and thus the recent 
star formation due to the dominating light of O and B stars \citep[e.g.][for the 
dependence of $\ml$ on the star formation timescale in different bands]{Into13}. 
Additionally, dust attenuation reduces the
observed light in the optical bands significantly at least in late type
galaxies and in the Milky Way. As is well known, the analysis of
galactic rotation curves suffers greatly from the uncertainty in $\mlv$ of the 
stellar disc. For extragalactic systems this
is usually done by adopting or fitting a reasonable $\ml$ value of the components. 
In a new approach \citet{Mar2013} used dynamic disc masses derived by a combination 
of integral field spectroscopy and a statistical scale height determination to 
derive $\mlk$ as a function of galactocentric radius for a set of galaxies.

With the increasing number of spatially resolved observations in the Near
Infrared bands (hereafter NIR), NIR luminosities are increasingly used for disc
mass determinations, primarily in the K band. The variation of $\mlk$
for different stellar populations is much smaller
than in the optical bands and
the extinction is smaller by a factor of 10 compared to the $V$ band.
Recently near- and far- infrared observations are used to determine disc masses of 
extragalactic systems \citep[see][and references therein]{Mar2013,McG14}. But there 
is still no direct method to determine for the same galaxy the surface mass density 
and the surface brightness independently. 
Another particular purpose of using $\mlk$ is to construct models of the Milky Way's
structure as constrained by star counts in the NIR in various Galactic
fields and NIR luminosity functions have been measured in many studies, e.g.
\citet{Jon87,Rue91,Wai92,Lop02,Pic03} to probe the Galaxy's structure. 
Star count methods are more tractable in
the NIR because of the greatly reduced extinction compared to optical. A local
K\,band LF for main sequence stars can also be obtained by converting the optically 
determined LF to the NIR using $V\,-\,K$ colours 
averaged over magnitude bins, as in \citet{MaSo82}, but the total brightness 
in the K\,band is dominated by giants.

It is expected that observations in the NIR are better suited 
than those in the optical bands to track the
distribution of stars, especially to
probe directly the mass-carrying population of late-type main sequence stars
(G, K, and M dwarfs). Since the solar neighbourhood is the only place where we
can determine directly both the luminosity and the stellar mass density, it is
worthwhile to investigate the properties of $\mlk$ based on the best available
data for this part of the Galaxy.

In this paper we present a fresh `ab-initio' determination of the Milky Way
LF in the NIR (we use the $K_\mathrm{s}$ filter, 2.2\,$\mu$m, throughout the
paper but skip the index 's' for clarity). Our study is primarily based on
{\it Hipparcos} stars and at the faint end on NIR data to be included in the
up-coming Fifth Edition of the Catalogue of Nearby Stars
(CNS5; Just et al., in preparation). 
These data have been obtained by identifying the CNS5
stars in the Two Micron All Sky Survey \citep[2MASS][]{skro06}.  The data set
drawn from the CNS5 is augmented by samples of stars selected from the
{\it Hipparcos} catalogue, for which 2MASS data is also available.

Our paper is organized as follows. In section \ref{datsamp} we describe the
construction of our data set and discuss its statistical completeness. In
section \ref{lumfunc} we derive the K\,band LF and investigate its
implications. We determine the contributions by the stars in the various
absolute magnitude bins to the luminosity and mass budgets of the Milky Way
disc in the local volume and in the solar cylinder, the principal result being
that even in the NIR the light is dominated by early type main sequence stars
and late giants. The consequences of this finding and our conclusions are
summarized in the last section (\ref{conc}).

%________________________________________________________________  Section 2

\section{Data sample}\label{datsamp}

\begin{figure}
  \centering
  \includegraphics[width=.98\hsize]{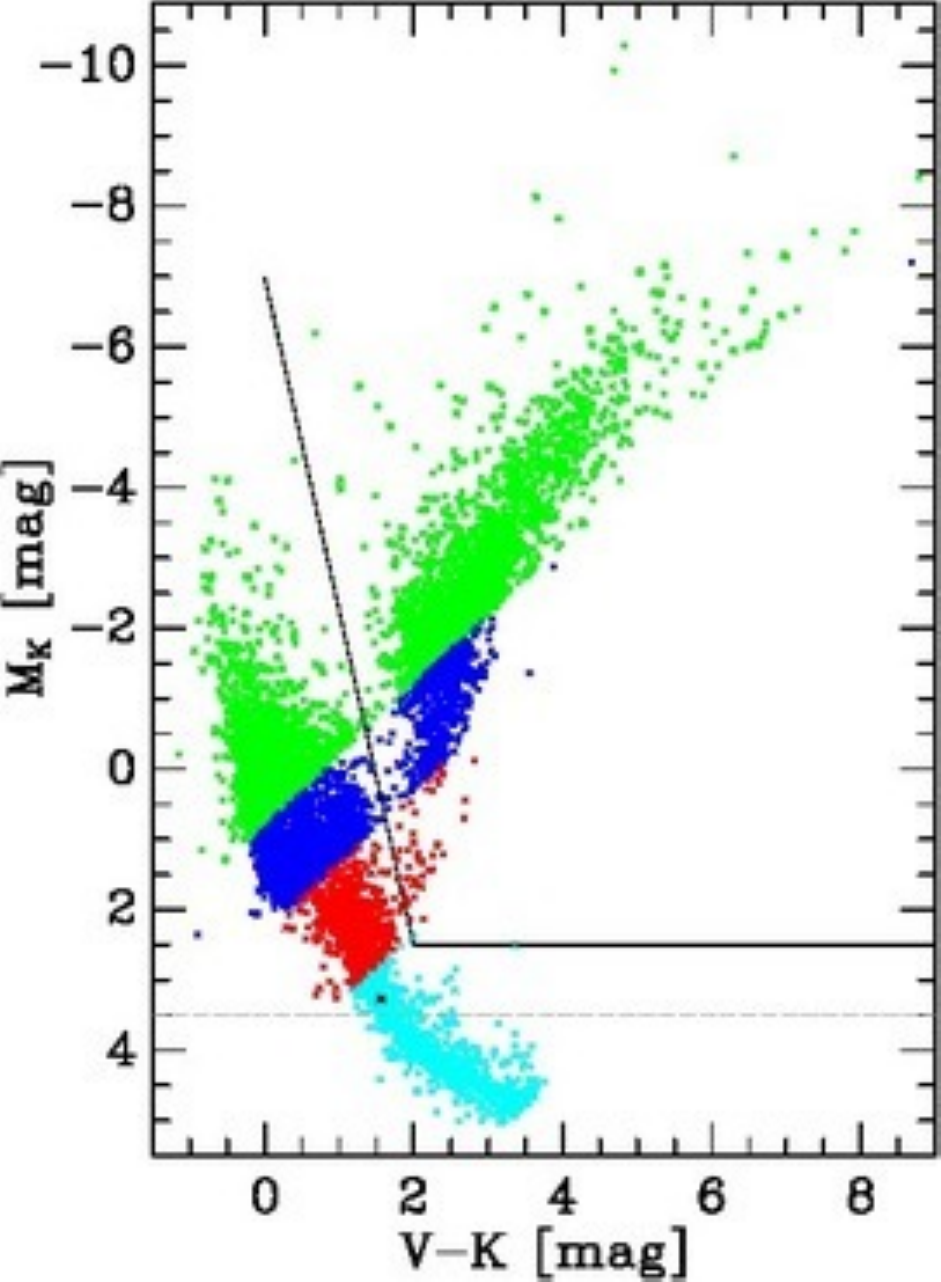}
  \caption{Colour-magnitude diagram of the {\it Hipparcos} stars used for this study.
    The colour coding is as follows -- green: stars with $M_V \leq$ 0.8\,mag
    and within 200 pc of the Sun; dark blue: stars with 0.8 $< M_V \leq$
    2.3\,mag and within 100\,pc; red: stars with 2.3 $< M_V \leq$ 4.3\,mag
    and within 40\,pc, and light blue: stars with 4.3 $< M_V \leq$ 8.3\,mag
    and within 25\,pc, respectively, see also table \ref{tab1}.
    In the construction of our LF we have used {\it Hipparcos} stars 
    with $M_{K} \leq$ 3.5\,mag (dashed line). The solid line cuts our giant sample
    off our dwarf stars. The black cross in the light blue dotted area denotes the Sun.
          }
  \label{fig1}
\end{figure}

We have constructed our stellar sample by merging two data subsets, in order to probe
the bright and faint ends of the LF respectively.
For the bright end of the LF we have extracted samples of stars from the
revised {\it Hipparcos} catalogue \citep{vanLeu}, using criteria of absolute
magnitude $M_V$ and the parallax limits summarized in columns 1 and 2 of Table
\ref{tab1}. The first three subsamples fall into the survey
part of the {\it Hipparcos} catalogue and are thus volume complete by construction. 
For the last subgroup ($d \leq$ 25\,pc), 
we sampled to apparent magnitude $V = 10.3$\,mag, 
based on the determination by \citet{JaWie97} that the {\it Hipparcos} catalogue
is complete down to absolute magnitude $M_V$ = 8.3\,mag. The numbers of stars in
each subgroup are tabulated in the fourth column of Table \ref{tab1}. Stars
with relative parallax errors larger than 15 per cent were excluded 
(except Antares, see below). The numbers of removed
stars are also given in Table \ref{tab1}. The vast
majority of all the sample stars were then matched to sources in the 2MASS
catalogue and their absolute $M_K$ magnitudes and $V\,-\,K$ colour were derived.
Only 10 stars did not appear in 2MASS: for these stars, $K$\,magnitudes were found in 
the literature or estimated from their known spectral types using the relation of
\citet{Ko83}. Since the nominal errors for the brightest stars in 2MASS are of
the order of 0.2\,mag, we have compared the 2MASS magnitudes with literature
values. We find that the differences between 2MASS and literature values are so
small that they can be ignored for our purposes. The resulting sample of
{\it Hipparcos} stars  is illustrated as
a colour-magnitude diagram (CMD) in Fig.~\ref{fig1}. It is complete 
to $M_K$ $\leq$ 4\,mag with respect to the $V$ band volumes and will be used down 
to the $M_K = 3 \pm 0.5$\,mag bin. Note that the discontinuities in
the CMD reflect Poisson statistics, because the subgroups
cover significantly different volumes. For the analysis we split the stars brighter 
than $M_K=2.5$\,mag into giants and dwarfs by the dividing 
line $M_K = 4.75(V-K) -7$ (see Fig.~\ref{fig1}). The resulting sample sizes are also 
given in Table \ref{tab1}.

The removal of stars with poor parallaxes (i.e.\ parallax errors $>15$ per cent)
means that we lose some stars which could significantly contribute to the total
luminosity, in particular at the sparse, bright end of the LF. The brightest
among the excluded stars are the M1.5\,Iab supergiant 
$\alpha$ Sco A `Antares' (parallax $5.89 \pm 1.00$\,mas, $M_K = -9.93 \pm 0.37$\,mag),
the M3\,III giant Hip\,12086 ($5.16 \pm 0.81$\,mas, $M_K = -5.72 \pm
0.34$\,mag), and the K0\,III giant  Hip\,61418 ($7.24 \pm 2.74$\,mas, $M_K =
-3.28 \pm 0.82$\,mag). We add Antares to our sample, because its inclusion corresponds 
to a factor of two in the star number in the $M_K = -10$\,mag bin 
and it adds 7 per cent to the total local 
luminosity. The contribution of all other stars is below 2 per cent and is 
therefore not taken into account. The case of Antares also shows, that the sampling 
of the bright end of the LF has large uncertainties despite the large volume 
with 200\,pc radius.

%__ _ _ _ _ _ _ _ _ _ __ _ __ _ _ _ _ _ _ __ [label 1] __  _ __ begin Table 1
\begin{table}
%  \begin{minipage}[t]{\columnwidth}
    \caption
        {Volume complete samples of {\it Hipparcos} stars.
        }
        \label{tab1}      
        %\centering          
        \begin{center}
        \renewcommand{\footnoterule}{}      % to avoid a line before footnotes
        \begin{tabular}{c c c c c c}     % 6 columns 
%          \noalign{\smallskip}
          \noalign{\smallskip}
          \hline\hline       
          \noalign{\smallskip}
          $M_V$\,[mag]  &  d$^{\rm lim}$\,[pc]  &
	  $N_{\mathrm tot}$ & $N(0.15)$ & $N_{\mathrm g}$  & $N_{\mathrm d}$ \\
           \hline
%               &    &   & &  \\
%          \noalign{\smallskip}
          \noalign{\smallskip}
             $\leq$ 0.8  &  200  &  4560 & 104 &  2660 & 1796 \\ %Antares removed from col.5
%	  \noalign{\smallskip}
          \noalign{\smallskip}  
             ]0.8, 2.3]   &   100  &  2039   & 21 &  541 & 1477\\
%                 &    &  &  &  \\
%	  \noalign{\smallskip}
          \noalign{\smallskip}  
             ]2.3, 4.3]   &   40  &  707   & 3 & 45 & 659 \\
%                 &    &  &  &  \\
%	  \noalign{\smallskip}
          \noalign{\smallskip}  
             ]4.3, 8.3]   &   25  &  694   & 20 & 2 & 672 \\
%                 &    &  &  &  \\
                       \noalign{\smallskip}
          \hline\hline 
        \end{tabular}
        \end{center}

{\it Note.} Column 1 lists the absolute $V$-magnitude ranges, col. 2 the distance
limits for completeness,
col. 3 gives the total number of stars, col. 4 the number of excluded stars
due to the parallax criterion (i.e.\ the relative parallax error for the stars is
greater than 15 per cent, except Antares), and col. 5 and 6 the number of 
selected giant and dwarf stars (see also Fig.~\ref{fig1}).
%  \end{minipage}
\end{table}      
%__ _ _ _ _ _ _ _ _ _ __ _ __ _ _ _ _ _ _ __  __ __ __ __  _ __ __end Table 1

%_________________________________________________ begin Figure 1
%_____________________________________________________end Figure 1

Since we derive distances and also base our absolute 
magnitudes on {\it Hipparcos} parallaxes the associated error (which we assume to be 
Gaussian) can affect our sample selection. The skewed error distribution after 
inverting 
parallaxes to distances results in the so called {\it parallax bias} \citep{franc} 
leading to larger mean distances. Another effect of the parallax error arising from 
the volume limit of 
the sample, but of opposite sign, is the Trumpler-Weaver or Lutz-Kelker 
bias \citep{TrWea,LK73} which increases star counts as 
the error volume outside our limiting distance is larger than inside and more stars 
should scatter in. In the literature these biases are often corrected 
statistically but it was shown that the 
Lutz-Kelker correction depends strongly on the adopted spatial distribution of the 
sample (usually a homogeneous density is assumed) and the true absolute magnitude 
distribution \citep[see e.g.][]{Smith03}. The latter may be well defined for special 
stellar types like red clump giants or cepheids. In our case for the bright Hipparcos 
stars there is no simple way to calculate the Lutz-Kelker correction due to the wide 
spread in luminosities and stellar types. In order to assess the impact of these biases 
on our star counts we select all stars in the {\it Hipparcos} catalogue.
We sample the parallax of each star 100 times randomly as 
a Gaussian with the mean of the original star's observed parallax and the standard 
deviation of its associated error. Then we perform the same cuts as 
before, divide the counts by 100 and compare them to our sample. The 40\,pc sample 
decreases by 1 per cent, the 100\,pc sample increases by 2 per cent, and the 200\,pc 
sample decreases by 2 per cent. 
Relaxing the parallax error cuts yield similar corrections.
In effect the corrections are small and show no clear trend in star 
counts as well as in the luminosity function so that we can safely neglect them.

A more severe source of error is dust extinction close to the Galactic plane. Due to
missing 3D maps of the highly inhomogeneous dust distribution in the solar neighbourhood we 
cannot correct for this error and only give rough estimates on its magnitude. For that 
we use the analytic extinction model from \citet{RyJus} based on \citet{Ver98} 
which gives a good estimate for the mean extinction. It represents a slab of constant dust 
density up to a distance of 55\,pc from the midplane with a hole at the solar position representing the 
local bubble. The resulting extinction
vanishes for stars closer than 70\,pc (inside the local bubble) 
and at Galactic latitudes $|b|>52^o$. The extinction reaches its maximum 
of about $A_V=0.19$\,mag at the limiting distance of 200\,pc in the Galactic plane. Applying 
the same procedure as before but now also correcting for 
extinction nothing changes for the 40\,pc sample. For the 100\,pc and 200\,pc samples 
the star counts increase by 2.4 per cent and 6.6 per cent, respectively. Overall, 
neglecting extinction leads to a slightly underestimated luminosity function  
for bright stars with $M_V < 2.3$\,mag. This bias is stronger for dwarfs at the upper main sequence, 
because of their stronger concentration to the midplane. On the other hand the derived luminosity 
distribution and total brightness in the K band is dominated by giants, which have a larger 
scale height and therefore being less obscured by extinction. 

\begin{figure}
  \centering
  \includegraphics[width=.98\hsize]{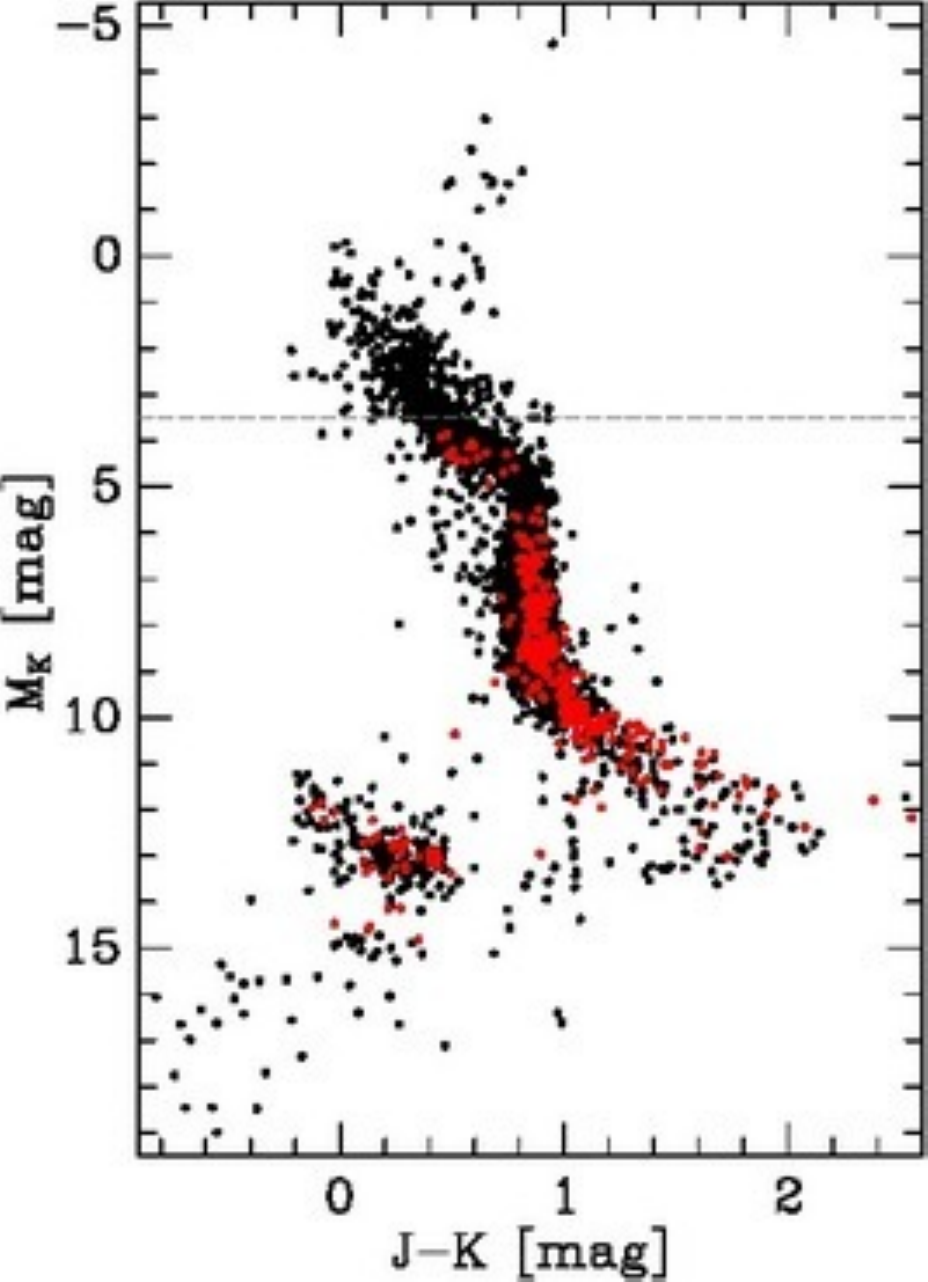}
  \caption{Colour-magnitude diagram representing our CNS5 stars. 
   Shown here are only stars with reliable distance estimates:  
   For 3520 stars (black) trigonometric 
   parallaxes are available, while 
   for the remaining 365 stars photometric distances were adopted (red symbols).
   In the actual construction of the LF we have used CNS5 stars fainter than
   $M_{K} =$ 3.5\,mag (dashed line).
           }
  \label{fig2}
\end{figure}

The faint end of the LF has been determined using an updated version of the 
Fourth Catalogue of Nearby stars \citep[hereafter CNS5, in prep.; CNS4,][]{JaWie97}. 
All CNS5 stars
within 25\,pc were cross matched with the 2MASS catalogue. 
From the 4622 CNS5 stars within 25 pc 135 stars got K-magnitudes from other
sources. Only 6 close binaries were removed at all as well as 9 brown dwarfs and
one white dwarf below the magnitude limit of the 2MASS survey.
For the missing stars,
$K$-magnitudes were obtained from the literature, from spectral types,
or applying an $M_{K}$--($V-K$) relation based on CNS5
stars with accurate parallaxes and reliable colours. The resulting CNS5 sample
is illustrated in Fig.~\ref{fig2} as a colour-magnitude
diagram $M_{K}$ vs ($J-K$).

%___________________________________________________ begin Figure 2
%_____________________________________________________ end Figure 2

%---------------------------------------------------------------subsection 2.1

\subsection{Completeness and vertical profiles}

\begin{figure}
  \includegraphics[width=.98\hsize]{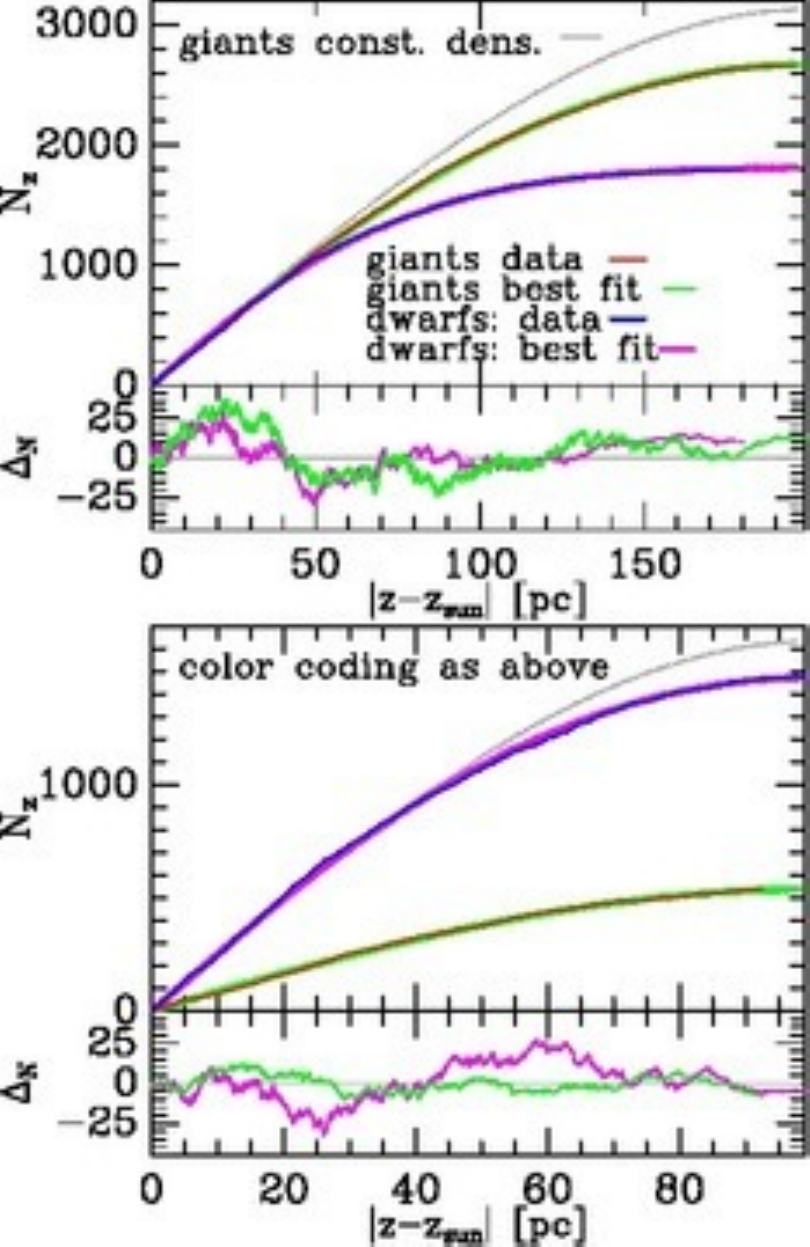}
  \caption{Top panel: Cumulative distributions $N_z$ in $|z-z_\mathrm{\sun}|$
    of the 2\,660 giants (red) and 1\,796 dwarfs (blue) in the
    200\,pc-sphere. Best fits
    of the vertical density profiles are shown in green and pink
    lines, respectively.  For comparison the grey dotted line shows the
    expectation for a constant density. The differences of model and 
    data $\Delta_N$ are also shown in the inset. The slopes at small $z$ of giants 
    and dwarfs determining the local number density differ by a few per cent only,
    which is by chance due to the sample selection in the 200\,pc sphere. The
    lower panel shows the same for the 100\,pc sample with 541 giants and 1477
    dwarfs.}
  \label{fig3n}
\end{figure}

Our {\it Hipparcos} samples are volume complete by construction. 
But beyond about\,$\pm 40$\,pc perpendicular to the Galactic plane, stellar
densities decline with increasing distance from the plane, and lead to
significant correction factors converting the mean luminosity density in the 
observed volume to the local luminosity density at the Galactic midplane. 
It can be easily shown that the impact of a 
(distance dependent) incompleteness, if present, on the cumulative
number of stars $N_\mathrm{z}$ as function of $z$ is mainly an apparently
reduced local density $n_0$, but the shape $N_\mathrm{z}/n_0$ is essentially
unaffected.  We thus use $N_\mathrm{z}$ to determine the vertical density
profiles and correct for the local volume density at the midplane
for the $r_0=$100\,pc and the 200\,pc samples, separately for both giant and
dwarf stars. Within a sphere of radius $r_0$, a constant density is obtained if
the cumulative number of stars $N_\mathrm{z}$ follows the relation
$N_\mathrm{z}=2\pi n_0 z(r_0^2-z^2 /3)$ (grey dotted lines in Fig.~\ref{fig3n}). 
We tested different vertical density
profiles (linear, exponential with and without a shallow core or Gaussian) and
investigated the impact of an offset of the solar position $z_\mathrm{\sun}$ from 
the midplane. It turns out that the profiles of the dwarfs in both
samples can be better fit with $z_\mathrm{\sun} = 15$\,pc.  

For the 200\,pc
sample, the cumulative star counts deviate significantly from the constant
density fit of the inner 80 and 50\,pc for giants and dwarfs, respectively,
which yields a local number density of $8.8 \times 10^{-5}\mathrm{pc}^{-3}$ for the
giants and 1 per cent less for the dwarfs.
Instead the giants are best fit by an exponential profile with a local number density
$n_0 = 9.35 \times 10^{-5}\mathrm{pc}^{-3}$ and an exponential scale height of
456\,pc. The
dwarfs are better fit by a cored exponential profile with flat density at the
midplane and local density of $n_0 = 9.90 \times 10^{-5}\mathrm{pc}^{-3}$ and an
exponential scale height of $z_{exp}=44$\,pc at large $|z|\gg z_{exp}$.
The flat profile near the midplane and the small scale height as well as the 
corresponding half-thickness
of $h=88$\,pc are expected for B and early A stars with velocity 
dispersions $\sigma\approx 5$\,km\,s$^{-1}$ (see Fig.~\ref{fig9}). The result is 
also consistent with the vertical density profile of the A star population 
derived in \citet{HoFl00}.
We note that the midplane densities are 5 - 10 per cent larger than
the value from the linear fit at small $|z|$.  The conversion 
factors $f_{200}=n_0/\langle n \rangle$ from the mean number 
densities $\langle n \rangle = N_{\mathrm fin}/V_{200}$ ($V_{200}$ is the volume of 
the 200\,pc sphere) to the local number densities $n_0$ are $f_{200,g} = 1.18$ 
and $f_{200,d} = 1.85$ for giants and dwarfs, respectively.

Investigating the 100\,pc sample
yields an almost constant density for the giants. The best fit with an exponential 
density profile yields a local
density $n_0 = 1.38 \times 10^{-4}\mathrm{pc}^{-3}$ and an exponential scale
height of 500\,pc (see lower panel of Fig.~\ref{fig3n}). The dwarfs of this
sample are best fit by an exponential profile with a local number density
$n_0 = 4.34 \times 10^{-4}\mathrm{pc}^{-3}$ and an exponential scale height of 173\,pc. 
The corresponding conversion factors are $f_{100,g} = 1.07$ and $f_{100,d} = 1.23$ for 
giants and dwarfs, respectively.

%_________________________________________________ begin Figure 3
%_________________________________________________ end Figure 3

\begin{figure}
  \centering
  \includegraphics[width=.98\hsize]{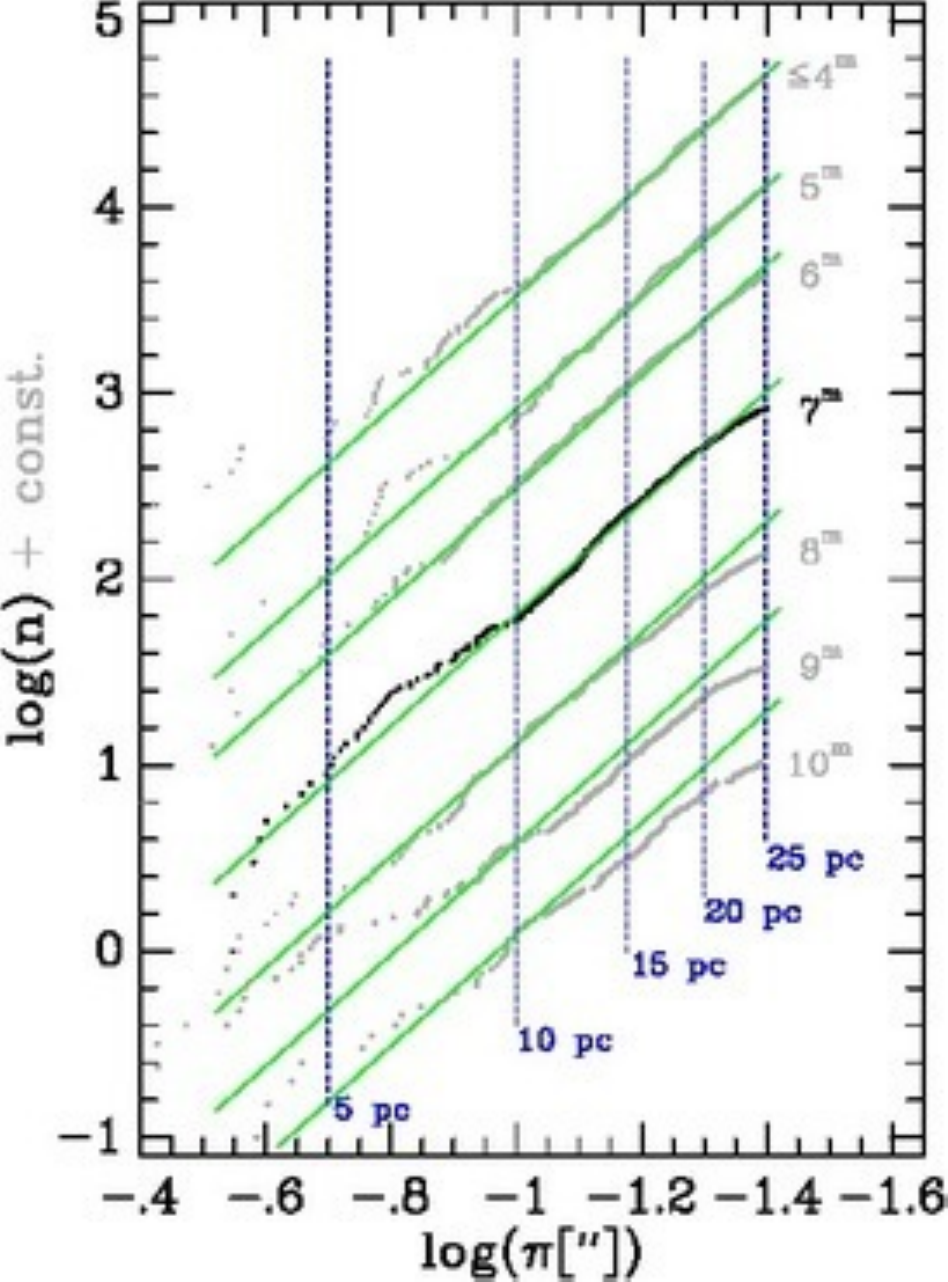}
  \caption{Radial cumulative star counts of CNS5 stars in magnitude bins (bin
    width 1\,mag) ranging from $M_K=4$\,mag and brighter to $M_K=10$\,mag (grey
    points).
    For clarity log(n) has been shifted arbitrarily along the vertical axis for all
    but the $M_K = 7$\,mag sample which is plotted in black.   
    The green lines indicate spatially homogeneous samples with the adopted 
    local density.}
  \label{fig3}
\end{figure}

Our faint star
counts are based on the CNS5 and suffer from incompleteness, which we
assume to depend on distance only. We have assessed the completeness of the CNS5 
by carrying out radial cumulative
star counts in each magnitude bin (with a width of 1\,mag). These are shown in
Fig.~\ref{fig3}. In a spatial homogeneous sample the cumulative number of stars\,$n$
grows with distance\,$d$ as $n\,\propto\,d^3$. In a double logarithmic log($n$) vs
($-$log($\pi$)) representation of
Fig.~\ref{fig3}  -- where $\pi$ is the stellar parallax --
a homogeneous sample would appear as a straight line with 
slope\,3 (green lines).

%___________________________________________________ begin Figure 4
%__________________________________________________ end Figure 4

The completeness limit of the CNS5 in each magnitude bin can be seen by a
deviation of the actual star counts from the line of slope 3. From
Fig.~\ref{fig3} we read off the completeness limits summarized in Table
\ref{tab2}. The magnitude bins $M_K$ = 13 and  14\,mag are dominated by white 
dwarfs. For magnitudes fainter than  $M_K$ = 14\,mag the completeness limit cannot 
be reliably determined. In order to derive a lower limit for the
star number densities we have considered a volume with a radius of 10\,pc. 
White dwarfs in the solar neighbourhood are completely sampled out to a distance of
$d \sim 13$\,pc from the Sun \citep{Holb}.  The stellar number
densities given in the next section have been determined within each
completeness limit and were then converted to a standard (spherical) volume
with a radius of 20\,pc.

\begin{figure}
  \centering
  \includegraphics[width=.98\hsize]{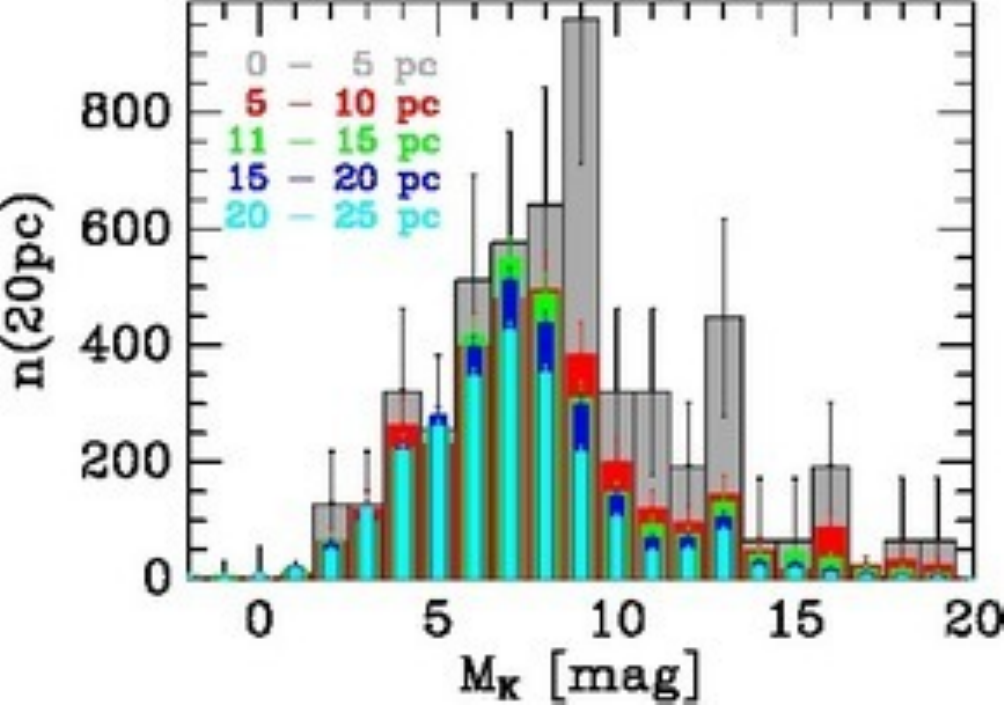}
  \caption{Illustration of the distance effect on 
   the NIR LF. The volume of the CNS5
   has been split into spherical shells of 5\,pc width.
   The LFs derived from each shell are over--plotted in colour coding onto
   each other.
           }
  \label{fig4}
\end{figure}

The luminosity function of the CNS5 stars alone is illustrated in
Fig.~\ref{fig4}. In order to demonstrate how the volume completeness of the
CNS5 influences the predictions of the LF, we have split up the original volume
of the CNS5 (with a radius of 25\,pc) into spherical shells of 5\,pc width. The
luminosity functions derived from each shell are over-plotted onto each other
in Fig.~\ref{fig4}. As can be seen here all shells give within
statistical errors consistent results up to an absolute magnitude of
$M_K$\,=\,5\,mag. Beyond that magnitude the outer shells yield reduced star
numbers, because they become increasingly incomplete. 

%______________________________________________________ begin Figure 5
%______________________________________________________ end Figure 5

%----------------------------------------------------------------- section 3
\section{Luminosity function and its Implications}\label{lumfunc}

We constructed
the local NIR luminosity function $\Phi(M_K)$ in terms of star numbers in 
the 20\,pc sphere $V_{20}=33\,510$\,pc$^3$ as described in
the previous section by combining the samples of stars brighter than
$M_K=3.5$\,mag drawn from the {\it Hipparcos} catalogue and fainter stars from the
CNS5. We will now discuss $\Phi(M_K$), 
the luminosity budget, and the mass functions that result. 

%---------------------------------------------------------------subsection 3.1
\subsection{Local luminosity function}

We show the results of our determination of the local luminosity 
function in Table \ref{tab2}. Given errors indicate the
(usually dominating) Poisson errors. We first note how smoothly the 
{\it Hipparcos} and CNS5 samples join together in the $M_K\,$=\,2\,and\,3 magnitude 
bins, see columns 3 and 5. The 
{\it Hipparcos} sample is volume complete down to $M_K\,$=\,4\,mag.  However, binary
stars have been treated more carefully in the CNS5 than in the {\it Hipparcos} sample
as evinced by slightly larger star numbers in these magnitude bins (cf.~Table
\ref{tab2}).
In column 9 of Table \ref{tab2} the luminosity function of the dwarfs alone (main
sequence and turnoff stars) is shown. This was derived by excluding giants 
according to the dividing line in the CMD as discussed earlier
(cf.~Fig.~\ref{fig1}) and similarly white dwarfs in the bins $M_K =$11--14\,mag
were removed.

%___  __ ___ _ _ ___ __ _ _ __ __ ___ __ _ ___[label 2]__ __ _ _ __ _ begin Table 2
\begin{table*}
  \begin{minipage}[c]{\textwidth}
    \caption
        {Stellar luminosity function $\Phi(M_K)$.}
        \label{tab2}      
        %\centering          
        \renewcommand{\footnoterule}{}      % to avoid a line before footnotes
        \begin{tabular}{c c c c c c c c c c}     % 10 columns 
          \noalign{\smallskip}
          \noalign{\smallskip}
          \hline\hline       
          \noalign{\smallskip}
%  $  $  $  $ HIP $ CNS $ comb $ MS $   $  $ \\
          \noalign{\smallskip}
   $M_K$         & Sp  & $\Phi_{Hip}$ & limits & $\Phi_{CNS}$ & err & 
   $\Phi_{comb}$ & err & $\Phi_{MS}$ & Mass \\
          \noalign{\smallskip}
   [mag] & [ ] & (*) & [pc] & (*) & (*) & (*) & (*) &
                 (*) & [${\mathcal{M}}_\odot$]   \\
          \noalign{\smallskip}
          \noalign{\smallskip}
     (1) & (2) & (3) & (4) & (5) & (6) & (7) & (8) & (9) & (10) \\
          \hline                    
          \noalign{\smallskip}
 $-$10&{M3,4 I--II}          &0.0024&  &         &    &   0.0024&0.0017&         &     \\
 $-$9 &{K--M2 I--II}         &0.0012&  &         &    &   0.0012&0.0012&         &     \\
 $-$8 &{M8 III}              &0.0145&  &         &    &   0.0145&0.0090&         &     \\
 $-$7 &{A,G I--II, M6,7 III} &0.0333&  &         &    &   0.0333&0.0101&         &     \\
 $-$6 &{M4,5 III }           &0.0767&  &         &    &   0.0767&0.0095&         &     \\
 $-$5 &{M1--3 III}           &0.186 &  &   0.5   &0.9 &   0.186 &0.015 &         &     \\
 $-$4 &{M0 III}              &0.321 &  &   0.    &    &   0.321 &0.020 &   0.0093&19.6 \\
 $-$3 &{B0,1, K4,5 III}      &0.699 &  &   0.5   &0.9 &   0.699 &0.030 &   0.0389&15.0 \\
 $-$2 &{K2,3 III}            &2.611 &  &   4.1   &1.5 &   2.611 &0.098 &   0.0981&10.8 \\
 $-$1 &{B2,3,G8--K0 III}     &3.492 &  &   1.0   &0.8 &   3.492 &0.157 &   0.531 & 6.1 \\
    0 &{B5--A0, F8--G2 III}  &5.512 &  &   7.7   &2.0 &   5.512 &0.441 &   3.135 & 3.46\\
    1 &{A2--A5}              &18.49 &  &   16.9  &3.0 &   18.49 &1.01  &   15.73 & 1.98\\
    2 &{F5}                  &53.95 &  &   48.1  &5.0 &   53.95 &2.57  &   51.94 & 1.47\\
    3 &{F8--G2 V}            &112.2 &  &   119.8 &7.9 &   112.2 &6.8   &   111.7 & 1.14\\
    4 &{G5--K4 V }           &184.3 &25&   220.2 &10.7&   220.2 &10.7  &   220.2 & 0.80\\
    5 &{K5--M1 V }           &76.80 &25&   261.1 &11.6&   261.1 &11.6  &   261.1 & 0.63\\
    6 &{M2--3 V }            &      &20&   397.0 &20.0&   397.0 &20.0  &   397.0 & 0.48\\
    7 &{M4--5 V }            &      &20&   512.0 &22.7&   512.0 &22.7  &   512.0 & 0.27\\
    8 &                      &      &10&   496.0 &63.0&   496.0 &63.0  &   496.0 & 0.16\\
    9 &                      &      &10&   384.0 &55.5&   384.0 &55.5  &   384.0 & 0.11\\
    10&                      &      &10&   200.0 &40.0&   200.0 &40.0  &   200.0 & 0.07\\
    11&                      &      &10&   120.0 &31.0&   120.0 &31.0  &   112.0 &     \\
    12&                      &      &10&    96.0 &27.8&    96.0 &27.8  &    80.0 &     \\
    13&                      &      &10&   144.0 &34.0&   144.0 &34.0  &    48.0 &     \\
    14&                      &      &10&    48.0 &19.6&    48.0 &19.6  &    16.0 &     \\
    15&                      &      &10&$>$ 16.0 &11.4&$>$ 16.0 &11.4  &$>$ 16.0 &     \\
    16&                      &      &10&$>$ 88.0 &26.6&$>$ 88.0 &26.6  &$>$ 88.0 &     \\
    17&                      &      &10&$>$ 24.0 &13.9&$>$ 24.0 &13.9  &$>$ 24.0 &     \\
    18&                      &      &10&$>$ 32.0 &16.0&$>$ 32.0 &16.0  &$>$ 32.0 &     \\
          \noalign{\smallskip}
          \hline \hline   
        \end{tabular}
\flushleft{Notes: Col 1: absolute $M_K$\,magnitude; 
  col 2: spectral type (after \citet{Jon87}); 
  col 3: luminosity function determined from the {\it Hipparcos} sample; 
  col 4: completeness limits of the CNS5 in abs. magnitude bins 4 to 10\,mag, radius 
         for upper limits at fainter magnitudes;
  col 5: luminosity function determined with the CNS5 stars; 
  col 6: errors to col.~5; 
  col 7: luminosity function combined from {\it Hipparcos} and CNS5;
  col 8: errors to col.~7; 
  col 9: resulting luminosity function for main sequence stars in the $K$\,band; 
  col 10: main sequence star masses.\\
  (*): The luminosity functions in cols. 3, 5, 7, and 9 as well as corresponding
  errors in cols. 6 and 8 are given in terms of stars per 20pc radius 
  sphere $V_{20}=33\,510$\,pc$^3$
  and per magnitude interval.}
  \end{minipage}
\end{table*}      
%_ __ ___ ___ ___ _ ___ _ __ __ ___ ___ __ _ ____ __ __ _ _ ___ end Table 2

\begin{figure}
  \centering
  \includegraphics[width=.98\hsize]{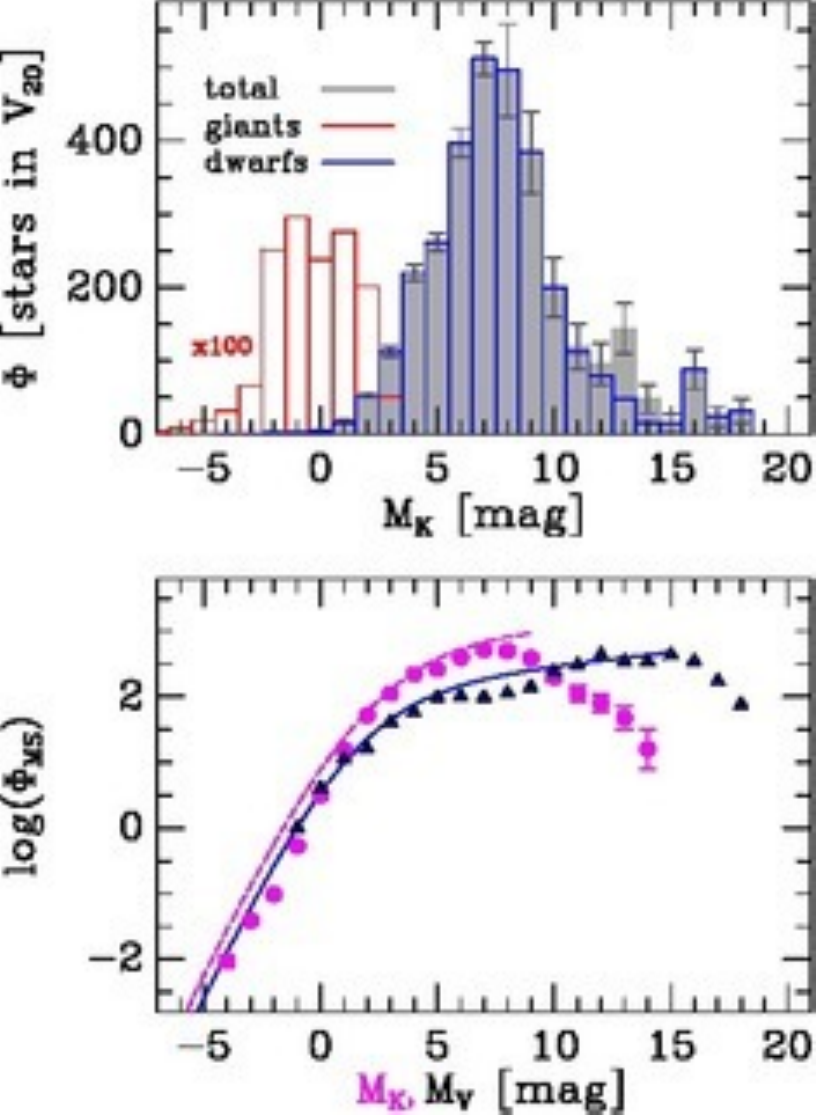}
  \caption{Top panel: Histogram of $\Phi(M_K)$ for all stars (grey histogram), 
    dwarf stars (blue) and
    giants (red). For clarity $\Phi(M_K)$
    of the giants has been enhanced by a factor of 100 in the red histogram. 
    The contribution of white dwarfs at $M_K \approx 13$\,mag is clearly visible.
    Bottom panel: Comparison of the main sequence 
    LFs in the V band (triangles) and in the NIR (pink circles).
    The blue line is an analytical fit to the optical data and the pink one
    a theoretical transformation of it to the K band. 
    $\Phi_{MS}$ is given as number of stars in the 20pc\,sphere $V_{20}$.} 
  \label{fig5}
\end{figure}

The total LF and the contributions of dwarfs and giants are shown in the top panel 
of Fig.~\ref{fig5}. Since the number of giants is very small, the giant LF 
is multiplied by a factor of 100 to make it visible. The dwarf LF shows a clear 
maximum at $M_K$\,=\,7--8\,mag beyond which it drops
by nearly an order of magnitude at $M_K$\,=\,13\,mag. The faint end of the LF in 
the L,\,T dwarf regime is unreliable due to the large error bars and incompleteness.
The shape of $log(\Phi(M_K))$ 
is roughly consistent with the LF observed in the optical bands as is
illustrated in Fig.~\ref{fig5}, bottom panel. 
Both the NIR and V band main-sequence LFs are
shown together with the analytical fit by \citet{MaSo82} to the V band LF of
\citet{Wie74}, and its theoretical transformation into the 2.2\,$\mu$m filter
band. As can be seen from Fig.~\ref{fig5} the analytic models fit very well 
in the F -- K dwarf regime.  At the very bright end ($M_K \leq 1$\,mag) the 
observed LF falls below the theoretical prediction.

%______________________________________________________ begin Figure 6
%______________________________________________________ end Figure 6

%-----------------------------------------------------subsection  3.2
\subsection{Local luminosity distribution}\label{lumdisb}

\begin{figure}
  \centering
  \includegraphics[angle=0, width=.98\hsize]{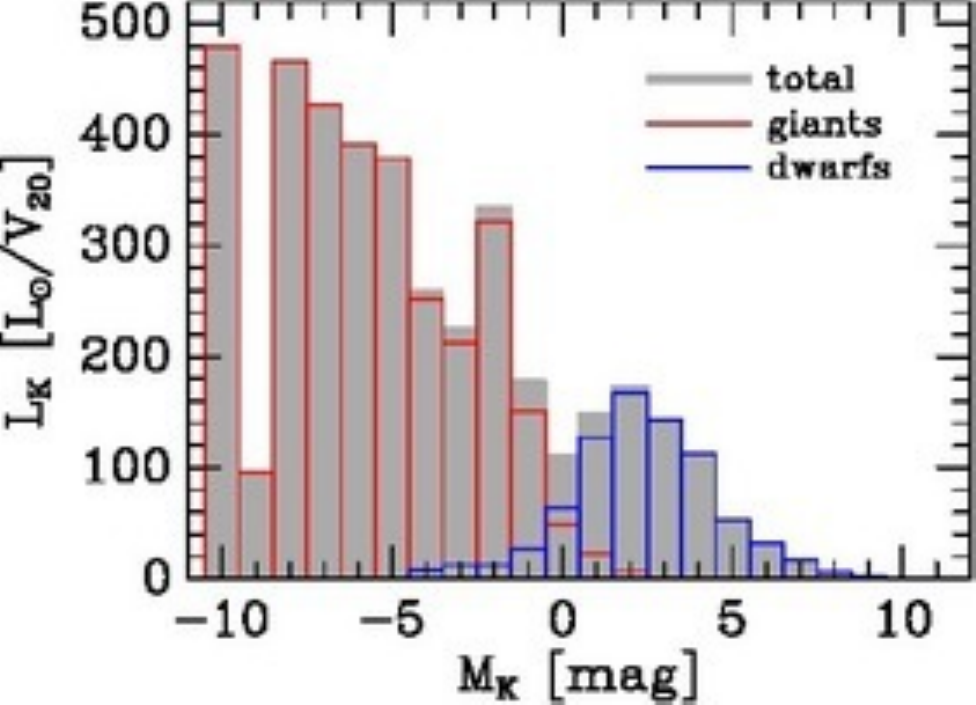}
  \caption{Local luminosity distribution in the NIR (grey histogram). 
           The contribution by giants is shown with the red line,  
           the blue line indicates the contribution by main sequence stars.
           }
  \label{fig6}
\end{figure}

In the previous section we have given the luminosity function $\Phi(M_K)$, 
i.e.\ the number of stars in absolute magnitude bins. In order to calculate 
the contribution of each bin to the total local luminosity in $V_{20}$, we now 
multiply $\Phi(M_K)$ with the mean luminosity
$L_K$\,=\,10$^{-0.4({M_K}-{M_{K\odot})}}L{_{K\odot}}$ of the bin, where $M_{K\odot}$
denotes the absolute $K$-magnitude of the Sun $M_{K\odot} = 3.27$\,mag \citep{Cas12}. 
The result is illustrated in Fig.~\ref{fig6} and again the contributions 
of dwarfs and giants are shown. The dwarf luminosity distribution peaks 
at $M_K = 2$\,mag corresponding to
F type stars and is analogous to the peak at $M_V = 1$\,mag in the
optical luminosity distribution \citep{Fly06}. The contribution of the faint end 
with $M_K > 8$\,mag is completely negligible. 
The luminosity distribution of giants does not show a clear maximum. 
In contrast, the contribution at the very bright end of supergiants 
beyond $M_K = -5$\,mag is dominating in the NIR. Due to the low number statistics 
at $M_K < -8$\,mag the contribution of the brightest end is very uncertain.
Red clump giants are responsible for the large value at $M_K = -2$\,mag.
In the V band the luminosity distribution of giants strongly declines 
for $M_V < 1$\,mag \citep{Fly06}. 

We find for the total K band luminosity 
$\rho_K=0.1205 \pm 0.0036\,L_{K\odot}$pc$^{-3}$, now converted to the local 
luminosity density. The uncertainty is dominated by the bright end of the giants. 
The giants dominate with 0.0971 $\pm$ 0.0036\,$L_{K\odot}$pc$^{-3}$ (80 per cent) 
and the dwarfs contribute 20 per cent 
with $0.0234 \pm 0.0005$\,$L_{K\odot}$pc$^{-3}$. 
A re--calculation of the V band luminosity of the same sample 
yields 0.053 \,$L_{V\odot}$pc$^{-3}$ \citep[slightly smaller than the old 
value $\rho_V=0.056\,L_{V\odot}$pc$^{-3}$ of][]{Fly06}. Combining the 
local K- and V band luminosity results in $(V-K) = 2.46$\,mag in the 
solar neighbourhood (with $(V-K)_\odot = 1.56$\,mag). This colour is slightly bluer 
than the value of $(V-K) = 2.55$\,mag determined in model A of \citet{JuJa10} 
based on a Scalo IMF.

%__________________________________________________ begin Figure 7
%___________________________________________________ end Figure 7

%----------------------------------------------------subsection 3.3
\subsection{Local mass distribution}\label{massdisb}

\begin{figure}
  \centering
  \includegraphics[width=.98\hsize]{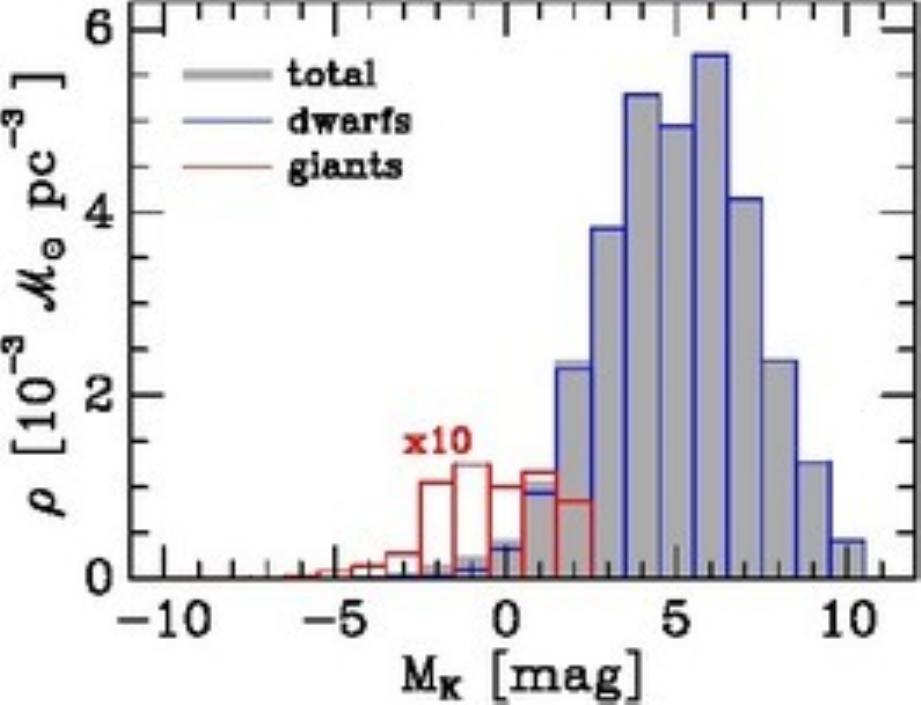}
  \caption{Distribution of the mass density $\rho$ for the stellar part of the Milky 
    Way disc (grey histogram). Stellar mass densities are given as a function of 
    absolute $M_K$ magnitudes
    (blue lines\,: main sequence stars, red lines\,: giants). For clarity the mass
    densities of the giants have been enhanced by a factor of 10 in the red
    histogram.}
  \label{fig8}
\end{figure}

We next examine how the NIR luminosity function relates to the mass
distribution of the stars. For this purpose we have multiplied the numbers of
stars with mean masses of the stars in each $M_K$ magnitude bin.  
For dwarfs, the
adopted masses are reproduced in the last column of Table \ref{tab2}. For
absolute magnitudes 3.1\,$\leq$\,$M_K$\,$\leq$\,9.8\,mag we have used directly
the absolute magnitude $M_K$-mass relation by \citet{HeCa}. For main sequence
stars brighter than $M_K=$\,3\,mag we have used the $M_V$-mass relation
compiled by \citet{And91} which we have transformed to $M_K$ by applying
$V-K$ main sequence colours. For red giants we assume a mass of 1.4 $\pm$ 0.1
${\mathcal{M}}_\odot$ as derived by \citet{Ste08} from asteroseismic
observations. Supergiants may be significantly more massive than that \citep{SchmiKa},
but are so few that their mass densities are negligible. 
The resulting mass distributions for giants and main sequence stars, which we 
have been able to estimate without
recourse to population synthesis modeling \citep{BelJo,Ziba}, are illustrated
in Fig.~\ref{fig8} as a function of absolute $M_K$ magnitude. The small 
contribution of giants is scaled up by a factor of 10 for visibility. In contrast to 
the luminosity distribution (see Fig. \ref{fig6}) the local mass density is 
dominated by the lower main sequence with $M_K>2$\,mag.

The local mass
density of main sequence and giant stars is $0.0315$ and 
$0.00058\,{\mathcal{M}}_\odot$ pc$^{-3}$, respectively. For the total stellar mass 
density we need to add the contribution of brown dwarfs and white dwarfs.
For brown dwarfs we add 0.002 ${\mathcal{M}}_\odot$ pc$^{-3}$ assuming a 50 per cent 
incompleteness in the observational data of late M, T and L dwarfs.
For the local mass density of white dwarfs we use 0.0032
$\pm$ 0.0003 ${\mathcal{M}}_\odot$ pc$^{-3}$  \citep{Holb}, which is similar to our 
finding of 0.0030 $\pm$ 0.0009 ${\mathcal{M}}_\odot$
pc$^{-3}$. This way, we find a total local mass density of 
$\rho = 0.0373\,{\mathcal{M}}_\odot$ pc$^{-3}$ for the stellar component.
The total stellar density is slightly smaller than in earlier determinations 
\citep[0.039, 0.044, 0.0415\,${\mathcal{M}}_\odot$ pc$^{-3}$;][respectively]
{JaWie97,HoFl00,Fly06}.
In all three publications a larger contribution of white dwarfs was adopted. 
Additionally, in \citet{HoFl00} and \citet{Fly06} the mass of the upper 
MS ($M_V<2.5$\,mag) was overestimated, and in \citet{HoFl00} the contribution of BDs 
was too high.

The local mass density implies a mass-to-light ratio of the stellar component in
the local volume of
$M/L_K\,=\,0.31\pm 0.02\,{\mathcal{M}}_\odot/L_{K\odot}$. This can be
compared with the optical mass-to-light ratio of 
$0.707\,{\mathcal{M}}_\odot/L_{V\odot}$, which is smaller than the 
value of $0.75\,{\mathcal{M}}_\odot/L_{V\odot}$ derived in
\citet{Fly06} mainly due to the smaller local mass density.

%_________________________________________________ begin Figure 8
%__________________________________________________ end Figure 8

%-----------------------------------------------------subsection 3.4
\subsection{NIR Surface brightness of the local disc}\label{surfbri}

\begin{figure}
  \centering
  \includegraphics[width=.98\hsize]{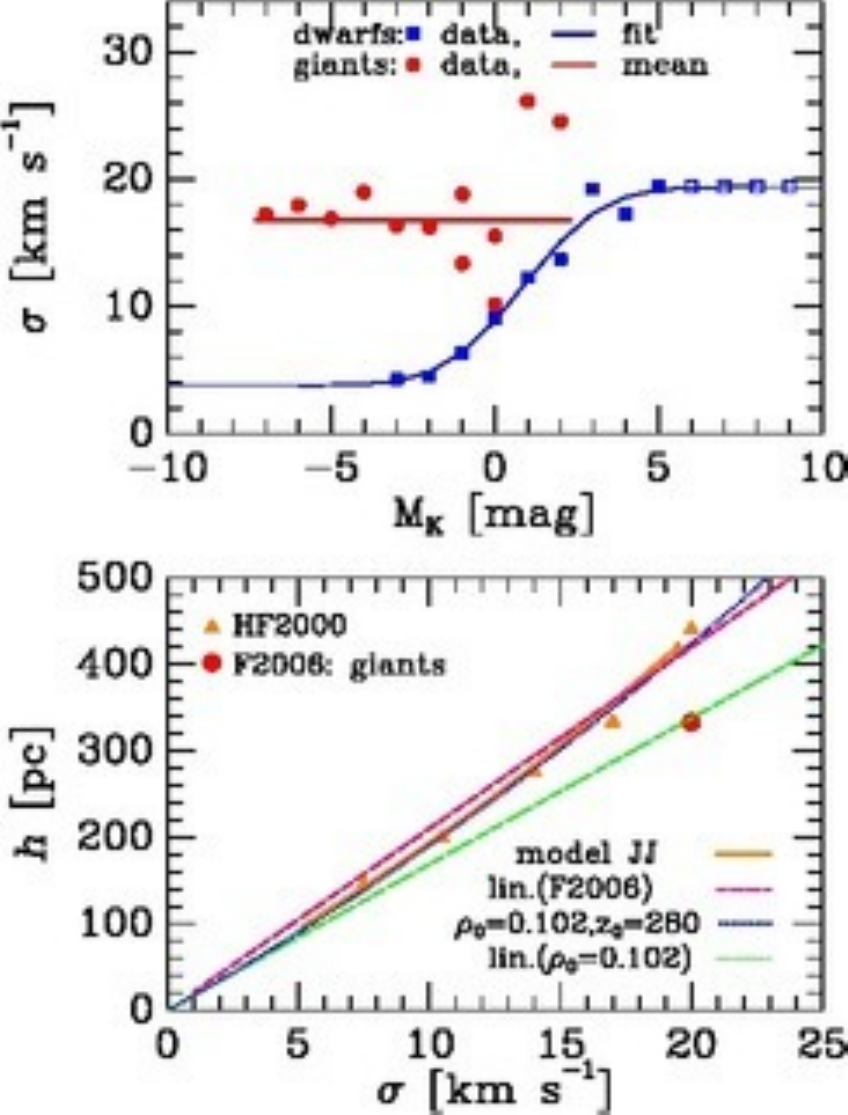}
  \caption{Conversion of velocity dispersion to thickness in 2 steps.  Top:
    Observed vertical velocity dispersions $\sigma$ 
    for dwarfs and giants in each magnitude bin
    together with the analytic approximations. For the meaning of open symbols,
    see text. 
    Bottom: Vertical thickness $h$ of the analytic fit
    used for the transformation (blue dotted line),
    for the best Just-Jahreiss-model (JJ model A, solid orange line), the values
    for the mass model used in \citet{HoFl00} (HF2000, orange triangles)
    and in \citet{Fly06} (red dashed line and F2006 for giants, red circle). 
    For comparison the linear
    extrapolation based on the total local density is also shown (green dashed
    line).}
  \label{fig9}
\end{figure}

The results presented so far reflect the NIR LF as well as the luminosity and mass
distributions as functions of $M_K$ for the Milky Way in the solar neighbourhood. 
More representative
for the entire Milky Way disc is the local surface brightness, i.e.\ the local
luminosity distribution multiplied by the vertical scale heights of the stars. 
The calculations of surface brightness and surface density depend on the adopted 
disc model, because for most sub--populations there are no direct observations of the 
vertical density profile available. It was
demonstrated in \citet{Fly06} that the vertical scale heights $h$ of the various
stellar populations can be roughly estimated by their vertical velocity
dispersions $\sigma$, because in a given gravitational potential the vertical scale
heights are in lowest order directly proportional to the latter.  For an
improved estimation of the surface brightness we use a higher order fit of the
(half-)thickness $h(\sigma)$ defined by $\Sigma=2\,h\,n_0$ connecting the local 
volume density $n_0$ and the surface density $\Sigma$ of the tracer. 
For a non--exponential vertical profile of the tracer the thickness $h$ differs 
from the exponential scale height.
Each stellar subpopulation is approximated by an isothermal component in the total 
potential characterized by the local 
density $\rho_0 = 0.102 {\mathcal{M}}_\odot$ pc$^{-3}$ and effective scale 
height $z_0=280$\,pc leading to
\begin{equation}
h(\sigma) = \sqrt{\frac{\pi}{2}}\cdot z(\sigma) + \frac{z(\sigma)^2}{3\,z_0} \quad\mbox{with}\quad
z(\sigma) = \frac{\sigma}{\sqrt{4\pi G\rho_0}}.
\end{equation}
The dotted line in the lower panel of Fig.~\ref{fig9} shows the approximation for 
the thin disc. The parameters $\rho_0$ and $z_0$ are chosen to reproduce the 
values $h(\sigma)$  of the detailed local
disc model of \citet{JuJa10} as well as the mass model used in
\citet{HoFl00,Fly06} (triangles in the lower panel of Fig.~\ref{fig9}).  
The parameters are optimized such that the fit function can also be used for the 
thick disc with velocity dispersion $\sigma\approx 40$\,km\,s$^{-1}$.

The upper panel of
Fig.~\ref{fig9} shows the measured vertical velocity dispersions in each $M_K$ bin and
{\it Hipparcos} group for the giants.  At the bright end the velocity dispersions
scatter around a constant value of 17.3\,km s${^{-1}}$, whereas the
faint giants in the 40\,pc group show a large scatter and a larger mean value of 
22.7\,km s${^{-1}}$. Since the outliers  do not contribute much to the total 
luminosity, we will use the same constant $\sigma_g=17.3$\,km s${^{-1}}$ for all 
giants. For the upper main sequence dwarfs in the {\it Hipparcos}
samples we show the mean values in each $M_K$ bin (weighted by the number
densities in the 20\,pc volume in order to avoid a bias due to the increasing
age with increasing $V\,-\,K$). 
To all lower main sequence stars we have assigned
the mean velocity dispersion of G,\,K dwarfs -- falling in the $M_K=5$\,mag bin --
from Table 4 of \citet{JaWie97} because the CNS5 is kinematically biased to
high proper motion stars at lower magnitudes and we expect 
-- and assume here -- the same kinematics for all these stars. Their $\sigma$ values
are shown as open symbols in Fig.~\ref{fig9}. 
In this figure the blue line shows the analytic fit, using a 
shifted error function, of
$\sigma(M_K)$ for the dwarfs, which we use to convert the velocity dispersions to the
corresponding thickness $h$.

%__________________________________________________ begin Figure 9
%____________________________________________________ end Figure 9

Additionally a correction for the thick disc contribution is
necessary. Adopting a standard isothermal old thick disc with a velocity
dispersion of 40\,km s$^{-1}$ corresponds to a thickness $h=1031$\,pc. In the 
solar neighbourhood we assume a thick disc fraction of 10 per cent for all giants 
and for the lower main sequence with
$M_K>1.5$\,mag. Due to the larger thickness this fraction is enhanced
accordingly for the surface brightness.  The surface brightness of each
subpopulation $i$ is given by $\Sigma_{K,i}=2\,h_{i}\,L_{K,i}$.  

\begin{figure}
  \centering
  \includegraphics[width=.98\hsize]{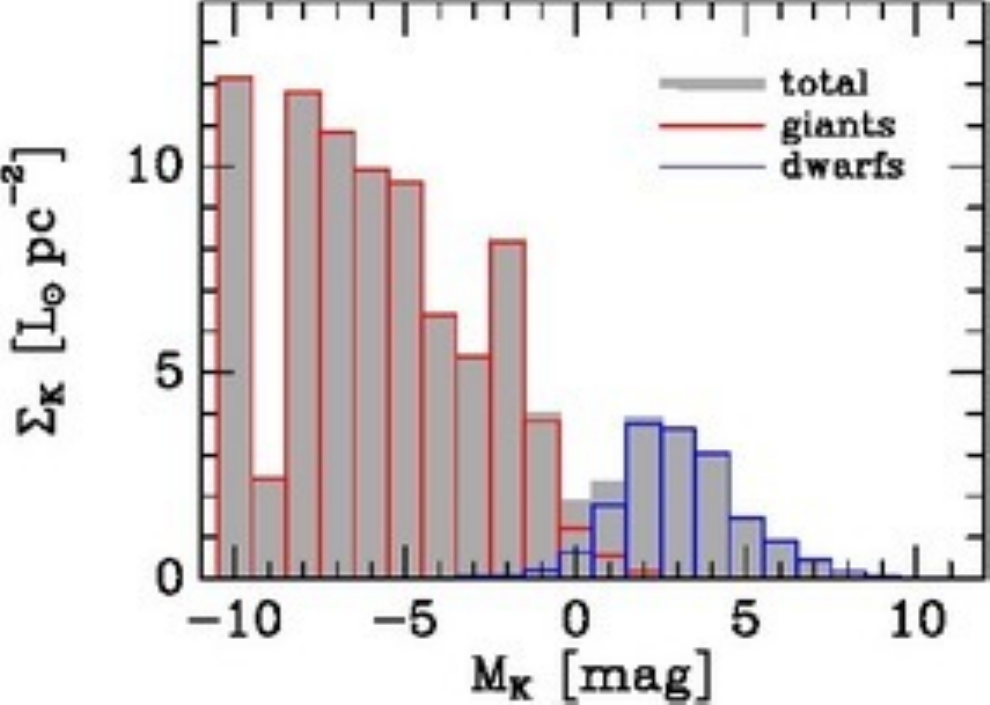}
  \caption{Surface brightness distribution in the NIR. 
           The blue and red lines indicate the contributions by the main 
	   sequence  stars and giants, respectively.
           }
  \label{fig7}
\end{figure}

The resulting surface
brightness distribution is shown in Fig.~\ref{fig7}, where we find similar 
features as in the local luminosity distribution (Fig.~\ref{fig6}). The distribution 
is strongly dominated by the bright end of the giants, the red clump giants are 
visible in the $M_K$\,=\, -2\,mag bin, and the dwarfs peak at the F dwarfs.
The integrated surface brightness is  
$\Sigma_{K}=98.7\,L_{K\odot}$pc$^{-2}=19.9$\,mag\,arcsec$^{-2}$ in total, composed by
$16.2\,L_{K\odot}$pc$^{-2}$ (16 per cent) for dwarfs and 
$82.5\,L_{K\odot}$pc$^{-2}$ (84 per cent)
for giants. 
Antares alone has added 5.3 $L_{K\odot}$pc$^{-2}$ demonstrating the uncertainty due 
to Poisson noise for the brightest supergiants. 

A similar determination of the V band surface brightness distribution shows a 
similar shape, but with a less dominant red giant contribution (cf.~discussion above). 
We find $29.1\,L_{V\odot}$pc$^{-2}=22.7$\,mag\,arcsec$^{-2}$. This value is 
19 per cent larger than the value of $24.4\,L_{V\odot}$pc$^{-2}$ determined 
by \citet{Fly06} arising from inconsistencies in the earlier transformation to 
surface brightness as can be seen 
by comparing their Figures 2, 5 and 6. In the I band, which \citet{Fly06} have used 
to derive the location of the Milky Way with respect to the Tully-Fisher (TF) 
relation, their Figures 9 and 10 seem to be correct.
Combining the 
K and V band surface brightnesses yields $(V-K)=2.89\,$mag for the solar cylinder. 

The corresponding stellar mass surface density derived from the local 
mass distribution (Fig.~\ref{fig8}) including white dwarfs and brown dwarfs 
is 33.3\,${\mathcal{M}}_\odot$ pc$^{-2}$, which is slightly smaller than the value 
35.5\,${\mathcal{M}}_\odot$ pc$^{-2}$ of \citet{HoFl04,Fly06}. It implies a K\,band 
mass-to-light ratio of $M/L_K\,=\,0.34\,{\mathcal{M}}_\odot/L_{K \odot}$. 
The corresponding  optical mass-to-light ratio is
$M/L_V\,=\,1.14\, {\mathcal{M}}_\odot /L_{V \odot}$.

%__________________________________________________ begin Figure 10
%____________________________________________________ end Figure 10

\subsection{Tully-Fisher relation}\label{tf}

\begin{figure}
  \centering
  \includegraphics[width=.98\hsize]{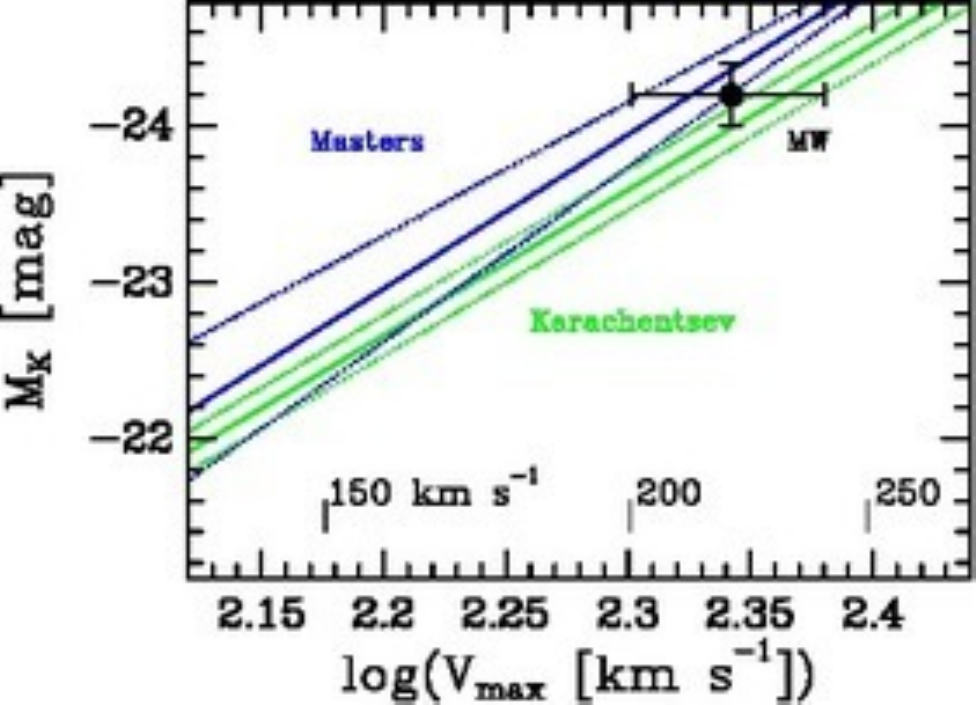}
  \caption{The Milky Way (dot with uncertainty bars, MW) with respect to the TF 
           relations 
           of \citet{Kar02,Mas14}, respectively. Dotted lines show the 1-sigma 
           scatter in the TFs.
           }
  \label{figTF}
\end{figure}

The surface brightness of the disc in the solar cylinder can be used to estimate the 
total K band luminosity of the Milky Way and compare it with the observed 
Tully-Fisher (TF) relation of extragalactic systems. We proceed similar as 
in \citet{Fly06} for the I band. The total disc luminosity is approximately independent 
of the adopted radial scalelength $h_R$ in the range of 2.5--5\,kpc. For definiteness 
we adopt $h_R=3.5$\,kpc and find with $\Sigma_{K}=98.7\,L_{K\odot}$pc$^{-2}$ at the 
solar radius of $R_0=8$\,kpc a total disc luminosity of $8.0\times 10^{10}\,L_{K\odot}$. 
Adding the bulge luminosity of $1.5\times 10^{10}\,L_{K\odot}$ 
\citep{Dri01,Por15} yields a 
total luminosity of $M_K=-24.2$\,mag. In Fig.~\ref{figTF} the data point adopting a 
maximum rotation velocity of the Milky Way disc of $220\pm 20$\,km\,s$^{-1}$ and 
an estimated uncertainty in the total luminosity of 0.2\,mag is shown. 
For comparison we plotted two determinations of the TF relation in the K band based on 
2MASS data.
\citet{Kar02} used edge-on\,galaxies to derive the isophotal TF relation in K band
using  $K_{20}=20$\,mag\,arcsec$^{-2}$. We have adapted their TF relation 
from Fig. 8 to a Hubble constant of $H_0=70$\,km\,s$^{-1}$\,Mpc$^{-1}$. 
For the Milky Way 
seen edge-on $K_{20}$ is at $R\approx20$\,kpc including more than 98 per cent 
of the total light. The Milky Way is 0.2\,mag brighter than the TF relation. For an 
alternative TF relation we show in Fig.~\ref{figTF} also the corrected equations 
of \citet{Mas08,Mas14} applied to an Sbc type galaxy. In this case the Milky Way 
is 0.2\,mag fainter than the TF relation. The systematic differences of the two 
TFs arise mainly on the inclination dependence of the $K_{20}$ isophotal 
luminosity \citep[see also][]{Said15} and the uncertainties in the extrapolation 
to the total luminosity. Deeper K band observations may solve this issue in the 
future. For the Milky Way seen face-on the isophotal luminosity would be 0.4\,mag 
fainter than seen edge-on. 

%__________________________________________________ begin Figure 11
%____________________________________________________ end Figure 11

%------------------------------------------------------------ section 4
\section{Conclusions}\label{conc}

We have constructed a new sample of stars representative for the solar
neighbourhood. The data set is based on two subsets separated by the intrinsic
brightness of the stars. The bright part comprises stars drawn from the survey
part of the {\it Hipparcos} catalogue. The faint part $M_K> 3.5$\,mag is 
based on the CNS5, an updated version of the Fourth
Catalogue of Nearby Stars. Each star has been identified in the 2MASS catalogue
so that we have individually determined NIR photometry in a homogeneous system
for all stars (typically $K_{\rm s}$) available.

For the Hipparcos stars we have selected volume complete samples 
in $M_V$ magnitude bins and corrected for the vertical density profile to determine 
the midplane density. These subsamples were then resampled in $M_K$ 
magnitude bins for $M_K \le 3.5$\,mag.  
The spatial completeness of the CNS5 sample has been carefully examined by
cumulative radial star counts for each $M_K$ bin resulting in a reasonable 
estimate of the local number density as faint as the $M_K=14$\,mag bin, 
which is well in the brown dwarf regime. For fainter stars we derived lower 
limits for the corresponding star numbers per magnitude bin. All star numbers
have been converted to a fiducial spherical volume with a
radius of 20\,pc (centered on the Sun) and constant density as a measure of the 
local volume density. For a detailed analysis we have separated the giants and 
white dwarfs from main sequence stars including turnoff stars and brown dwarfs. 
We have then determined the NIR luminosity function of the
stars in the Milky Way disc by direct star counts in our sample. The K band 
luminosity function shows a strong maximum at $M_K=7-8$\,mag. 
At $M_K=13-14$\,mag the white dwarfs dominate the star counts.

The luminosity
function has then been converted to the luminosity distribution of the stars by
multiplying the star numbers with the typical luminosities of the stars in each
absolute magnitude bin. The resulting (midplane) luminosity distribution is 
strongly dominated by the very bright end of giants and supergiants. 
A secondary peak at
$M_K$\,$\approx$\,2\,mag is due to A -- F type main sequence stars while 
at $M_K=-2$\,mag the red clump 
giants stick out. At the bright end there is no decline measurable, 
which means that the low number statistics of the brightest supergiants dominate 
the uncertainty of the total local luminosity in the K band. We find a total 
luminosity density of $\rho_K=0.121\pm0.004\,L_{K\odot}$pc$^{-3}$, where the giants 
dominate with a contribution of 80 per cent.
Combined with the V band luminosity density of $\rho_V=0.053\,L_{V\odot}$pc$^{-3}$ 
we find a value of $V-K=2.46$\,mag for the colour in the solar neighbourhood.

We have determined the mass
distribution of the stars as probed by the NIR luminosity function in the same way. 
Quite
contrary to the NIR light, the mass of the Milky Way is dominated by K and M main
sequence stars.  We conclude from this discussion that the mass--carrying
population of stars in galactic discs cannot be observed directly in the NIR 
on a star-by-star basis.
The total mass density of the stellar component 
is $\rho=0.0373\,M_{\odot}$pc$^{-3}$, which is about 10 percent smaller than earlier 
determinations due to reduced contributions by white dwarfs and brown dwarfs. 
The local mass-to-light ratio is 
then $M/L_K\,=\,0.31\,{\mathcal{M}}_\odot/L_{K \odot}$.
The corresponding corrected value in the optical 
is $M/L_V\,=\,0.71\,{\mathcal{M}}_\odot/L_{V \odot}$.

For a comparison to extragalactic systems it is important to determine the surface 
brightness of the disc. We have used a detailed vertical disc model to derive the 
effective thickness of the stellar populations in the magnitude bins and took into 
account a correction for the thick disc contribution with a larger thickness. 
The resulting surface brightness function shows similar features as the local 
luminosity distribution.
The total surface brightness 
is $99\,L_{K\odot}$pc$^{-2} = 5.4\times 10^{-6}$ W m$^{-2}$ with 84 per cent 
resulting from giants. This value can be compared to the
$K$ band surface brightness of
the disc determined from DIRBE data after removing all point sources yielding 
$68\,L_{K\odot}$pc$^{-2}$ \citep{Melchior}. The difference corresponds to the 
contribution of all supergiants with $M_K<-8$\,mag, which seems reasonable.
The mass model yields a surface density of $33.3\,M_{\odot}$pc$^{-2}$ and a
mass-to-light
ratio for the solar cylinder of $M/L_K\,=\,0.34\,{\mathcal{M}}_\odot/L_{K \odot}$. 
The corresponding  optical surface brightness and mass-to-light ratio 
is $29.1\,L_{V\odot}$pc$^{-2}$ and
$M/L_V\,=\,1.14\, {\mathcal{M}}_\odot /L_{V \odot}$, respectively. 
With the redetermination of the surface brightness we have corrected a bug in
the earlier determination by \citet{Fly06}, which happened particularly in the V band.
An extrapolation of the local surface brightness to the whole Milky Way yields a total 
K band luminosity of $M_K =-24.2$\,mag. With standard values of the disc properties the Milky 
Way falls between the \citet{Kar02} and \citet{Mas14}  K band TF relations.

The stellar population in the solar neighbourhood is strongly dominated by 
young stars compared to the population in the solar cylinder due to the much 
smaller scale heights of the young populations. Nevertheless, the mass-to-light 
ratio in the K band is only 10 per cent larger in the solar cylinder. 
The reason for the luminosity of the present day giants -- dominating 
the light in the K band -- being a rough measure of stellar mass, carried 
by F, G, and K stars of the lower main sequence, is the similarity of their 
age distributions. The birth time distribution for the precursors of the 
giants -- mainly F and G dwarfs -- is spread over the age of the disc 
and similar to that of the F, G, and K dwarfs still on the 
main sequence. As a consequence, the dynamical evolution is similar and
produces comparable scale heights. Thus we conclude 
that the mass-to-light ratio does not vary strongly in disc populations with a long 
star formation history and a calibration of the absolute value is provided by the 
solar neighbourhood properties.

The colours and mass-to-light ratios are consistent with the stellar populations 
derived in the local disc model of \citet{JuJa10}. 
\citet{Into13} determined mass-to-light ratios and colours for disc populations 
with exponentially declining star formation histories. Our values are roughly 
consistent with these models for relatively flat star formation histories. An 
extrapolation from the solar radius to the whole disc would shift the colour 
and mass-to-light ratios slightly dependent on the disc growth model. 
Our K band mass-to-light ratio of the solar cylinder 
of $0.34\,{\mathcal{M}}_\odot/L_{K \odot}$ is very close to the mean value 
of $0.31\,{\mathcal{M}}_\odot/L_{K \odot}$ of 30 disc galaxies 
derived by \citet{Mar2013}.

\section*{acknowledgements}
This work was supported by the Collaborative Research Centre SFB 881
'The Milky Way System' (subproject A6) of the German Research
Foundation (DFG). This publication makes use of data products from the 
Two Micron All Sky Survey,
which is a joint project of the University of Massachusetts and the Infrared
Processing and Analysis Center, funded by the National Aeronautics and Space
Administration and the National Science Foundation. This research has made use
of the SIMBAD and VIZIER databases, operated at CDS, Strasbourg, France.
 
We thank Laura Portinari for fruitful discussions accompanying this project.

{}

\end{document}